\documentclass[10pt,journal,compsoc]{IEEEtran}
\ifCLASSOPTIONcompsoc
  \usepackage[nocompress]{cite}
\else
  \usepackage{cite}
\fi
\usepackage{tikz, amsmath, amssymb, bm, color}
\usepackage{float}
\usepackage{multicol}
\usepackage{bbding} 
\usepackage[ruled, vlined, linesnumbered]{algorithm2e}
\usepackage{mathtools}
\usepackage{siunitx}
\usepackage{pifont}
\usepackage{paralist}
\usepackage{enumitem}
\usepackage{hyperref}

\usetikzlibrary{automata, positioning, arrows}

\newcommand*\circled[1]{\tikz[baseline=(char.base)]{
            \node[scale=1.0,shape=circle,draw,inner sep=3.5pt] (char) {};
            \node[scale=1.0] (char) {\textcolor{black}{\footnotesize{\textbf{#1}}}}}}

\newcommand*\blackcircled[1]{\tikz[baseline=(char.base)]{
            \node[scale=1.0,shape=circle,draw, fill,inner sep=3.5pt] (char) {};
            \node[scale=1.0] (char) {\textcolor{white}{\footnotesize{\textbf{#1}}}}}}

\SetKw{Continue}{continue}

\newcommand{\additionColor}{black}
\usepackage{etoolbox}

\begin{document}
\title{\color{\additionColor}ETVO: \textit{Effectively} Measuring Tactile Internet with Experimental Validation}
%
\author{\IEEEauthorblockN{H.J.C.~Kroep, V.~Gokhale, J.Verburg, R.~Venkatesha~Prasad\\}
\IEEEauthorblockA{Embedded and Networked Systems, Delft University of Technology, The Netherlands\\
}
\thanks{This work is a significant extension of our earlier work in \cite{verburg2020setting} published in IEEE INFOCOM 2020.}
\thanks{Author emails:\{H.J.C.Kroep, V.Gokhale, R.R.VenkateshaPrasad\}@tudelft.nl.}
}


\IEEEtitleabstractindextext{%
\begin{abstract}
The next frontier in communications is \textit{teleoperation} -- manipulation and control of remote environments with feedback. Compared to conventional networked applications, teleoperation poses widely different requirements, ultra-low latency (ULL) is primary. Realizing ULL communication demands significant redesign of conventional networking techniques, and the network infrastructure envisioned for achieving this is termed as \textit{Tactile Internet} (TI). The design of the network infrastructure and meaningful performance metrics are crucial for seamless TI communication. However, existing performance metrics fall severely short of comprehensively characterizing TI performance. We take the first step towards bridging this gap. We take Dynamic Time Warping(DTW) as the basis of our work and identify necessary changes for characterizing TI performance. Through substantial refinements to DTW, we design \textit{Effective Time- and Value-Offset (ETVO)} -- a new method for measuring the fine-grained performance of TI systems. 
{\color{\additionColor}Through an in-depth objective analysis, we demonstrate the improvements of ETVO over DTW. Through human-in-the-loop subjective experiments, we demonstrate how and why existing QoS and QoE methods fall short of estimating the TI session performance accurately. Using subjective experiments, we demonstrate the behavior of the proposed metrics, their ability to match theoretically derived performance, and finally their ability to reflect user satisfaction in a practical setting. The results are highly encouraging.}
\end{abstract}


%
}
\maketitle
%

\iffalse
\else
\section{Introduction}\label{sec:intro}
{\color{\additionColor}The Covid-19 pandemic has made us realize the power of the Internet yet again by seamlessly connecting persons located remotely through audio-video interaction. \textit{Tactile Internet (TI)} promises to advance this level of immersion by augmenting a new modality of interaction -- \textit{haptic} (touch), via force feedback. This enables us to perform teleoperation, a primary driving force for the realization of Industry 4.0 revolution~\cite{fettweis2014tactile,maier2016tactile}. 
One of the most well-known use cases of TI is telesurgery, where a patient in the controlled domain can be operated upon by a surgeon (\textit{operator}) in the master domain, from a distant location as effectively as in conventional surgery. The position and rotation of the surgeon's arm are replicated by a remote robot arm (\textit{teleoperator}), while simultaneously, audio-video and force measurements from the controlled domain are fed back to the surgeon. This \textit{transportation of skills} allows the surgeon to perform the medical procedure as if he/she is physically present in the controlled domain. A schematic representation of TI is shown in Figure~\ref{fig:ti-arch}.

Telesurgery is one of the examples of many potential applications of TI. Management of disaster-struck areas remotely, and telerepair and telemaintenance of machinery, and also other sectors like education, health, and manufacturing \cite{Aijaz2018} will benefit from TI. Deploying TI will significantly save time, money, lives, and in many cases, enable applications that are otherwise not feasible. A case in point is cleaning the Fukushima Daiichi disaster site without putting humans in harm's way, which otherwise would have exposed them to hazardous nuclear waste \cite{kawatsuma2012emergency}. Such complex applications cannot be performed by autonomous robots alone and thus necessitate human expertise. 

TI applications have one requirement in common -- force feedback. For instance, a surgeon must make a shallow cut in the skin without damaging the underlying tissues. Force feedback is critical to enable him/her to apply an optimal amount of pressure to control the depth of the cut. While performing tasks relying on force feedback, it is vital to minimize the roundtrip delay to ensure the real-time performance of tasks -- if a delay of the force feedback is high, the cut could be deep into the underlying tissue before the operator realizes, thus leading to a catastrophe. TI applications require ultra-low roundtrip delay guarantees.
%
%
\begin{figure}[t]
    \centering
    \includegraphics[width=\columnwidth]{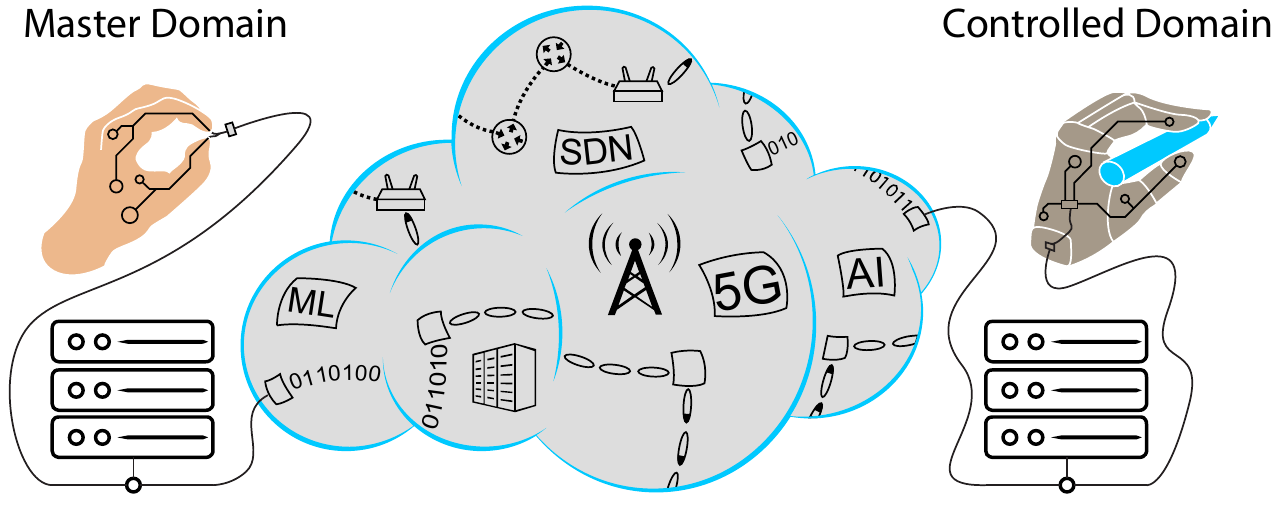}
    \caption{A schematic representation of Tactile Internet showing the master and controlled domains.}
    \label{fig:ti-arch}
    \vspace{-8pt}
\end{figure}
Both TI and conventional teleconferencing applications benefit from the low delay. However, there is a strong distinction in the way delay manifests. In teleconferencing applications, communicating with a high delay can be frustrating and inefficient, but it does not hamper one's ability to speak. On the contrary, in TI applications, delayed force feedback can significantly impair the operator's ability to teleoperate. 
Another commonly stated requirement for TI is ultra-high reliability. Both delay and reliability demands of TI are not supported on current networks.

TI applications have several layers of complexity. The first complexity is that incorrectly reproduced actions can result in inappropriate and often catastrophic outcomes based on the task environment. The second complexity is the presence of force-feedback. Force feedback heavily influences the operator's ability to perform tasks which is dependent on roundtrip delay. It affects the performance at the controlled domain, which in turn affects the observed feedback and spiraling into unbounded error and eventually breakdown of the applications. 
The third complexity is due to significantly higher dependence on a human operator. It is challenging to characterize the user experience and optimize the application (over delay and reliability) with humans adding high non-linearity in the loop. These complexities make it challenging to create objective metrics for measuring and comparing TI session performances.

Thanks to the vision and recent advancements in the field of 5G, TI will benefit enormously~\cite{budhiraja2019tactile,  Millnert2018, ge20195g, ali2021urllc} in the context of ultra-reliable, low-latency communication (URLLC). The path towards offering the necessary performance for TI applications has been advancing rapidly. However, several major challenges need to be addressed before we realize a functional TI. One of those challenges is the subject of this work: \textit{assessment of the quality of a TI session precisely and objectively}. This is crucial for several reasons:
\begin{inparaitem}
\item[\circled{1}] estimating the quality of a session \textit{a priori} is essential for executing mission-critical applications, \item[\circled{2}] adapting TI application parameters based on network dynamics, \item[\circled{3}] benchmarking novel solutions at various layers of the TI protocol stack. 
\end{inparaitem}

\vspace{1mm}
\noindent\textbf{Measuring TI performance.} The core task of a TI system (comprising sensors, actuators, communication, and computing entities) is to communicate haptic-audio-video modalities that are observed in the master and controlled domains. The TI system generally strives to match the signal as accurately as possible. An ideal system would recreate the values at the controlled domain identical to the sensed values -- in \textit{value/magnitude} and \textit{time} -- and vice versa.
However, in practice, such a performance is not feasible. The reconstructed signal can be degraded in time due to varying delay and in value because of losses.
For fine-grained performance analysis, it is crucial to distinguish offset caused in the two domains as independently as possible, which is a major challenge.

Literature provides two types of approaches for measuring TI performance -- Quality of Service (QoS) and Quality of Experience (QoE). 
The QoS metrics are based on standard network performance indicators such as delay, jitter, reliability, and throughput. In the case of TI, the emphasis is on the end-to-end delay, for which a commonly stated target is \SI{1}{ms} \cite{fettweis2014tactile,wollschlaeger2017future,maier2016tactile}. It should be noted that end-to-end refers to the entire TI application and includes sensing, interim computation, actuation, and the network. To start with, the target is to achieve around 15\,ms~\cite{holland2019} and the reliability in the order of around 99.999\%~\cite{fettweis2014tactile,wollschlaeger2017future,maier2016tactile}. 
While QoS metrics serve as indicators of the network dynamics, they fall short of completely characterizing the TI performance since they are oblivious to the underlying signal properties. In particular, it is challenging to evaluate signal-aware solutions and adapt their behavior to the underlying network. A well-known example of a signal-aware solution is the \textit{Perceptual Deadband (PD)} protocol~\cite{hinterseer2005novel,hinterseer2008perception}, which exploits the limits of human perception to reduce the data rate and is therefore not agnostic to the TI application. In general, QoS metrics have severe limitations in characterizing TI applications, unlike teleconferencing type of applications.

QoE is a human-centric approach, and it focuses on an objective evaluation of user experience. 
Ideally, a QoE metric should closely estimate user experience without involving an extensive user study where subjects interact through a TI system and subjectively grade their experience. Several algorithms have been designed to calculate an objective metric based on sensed and reconstructed signals. The problem with such approaches is that they cannot distinguish between degradation (offset) in time and value domains. For example, the works in~~\cite{Chaudhari2011,Yuan2014,hamam2013toward} propose RMSE-based metrics.
Such solutions do not consider delay and therefore face problems caused by a mismatch between two signals in time, as illustrated in Figure~\ref{fig:ti-blackbox}. As shown in the figure, the shape of the sensed signal needs to be faithfully reproduced even when packet losses and jitter are present. To this end, a major tool that has been used by many is Dynamic Time Warping (DTW)~\cite{muller2007dynamic} to find the similarity in signals. A well-known use of DTW is in speech recognition, where it can identify similarities in speech irrespective of pitch, speed, and amplitude. Further, the lack of a framework for the performance characterization of TI at a fine-grained level severely impedes efforts to measure QoE. This forms our primary motivation. 

}
\vspace{2mm}
\noindent\textbf{Our contributions.} Our approach is to devise a method that is capable of extracting fine-grained time- and value-offset between the sensed and reconstructed signals in a TI session. This method can be applied to any end-to-end TI system (starting from sensors on one end to the actuators on the other end) in a manner that is agnostic to the underlying network. 
We take DTW -- since it is the widely used tool for determining sample-wise similarity between the two time sequences -- as the starting point, show its limitations in the context of TI, and build on it. Our contributions in this work are as follows.

%
\begin{itemize}[leftmargin=*]
\item  \color{black}We present a detailed analysis on characterizing TI sessions with DTW and identify areas of improvement for the stated task (Section~\ref{sec:limitations}). 
\item We present a concrete mathematical framework, which we call \textit{Effective Time- and Value-Offset (ETVO)}, that extracts fine-grained time and value-offset between sensed and reconstructed signals of a TI system. To the best of our knowledge, this is the first of its kind work that comprehensively characterizes TI session performance in a system-agnostic manner. 
(Section~\ref{sec:etvo}). 
\item \color{\additionColor}We propose two novel metrics -- average effective time-offset ($T_\text{ETVO}$) and average effective value-offset ($E_\text{ETVO}$). These metrics can be jointly used to compare the performance of different TI solutions.
\item Through objective analysis using a realistic TI setup, we demonstrate the effectiveness of ETVO and its improvement over DTW  (Section~\ref{sec:setup}).
\item 
To validate ETVO, we conduct human subjective experiments on a realistic TI setup under a wide variety of network settings. We show that the proposed metrics correlate well with the user grades (Section~\ref{sec:subjective}).
\item Independently, we theoretically derive the expected average delay of the TI sessions and show that it corroborates well with $T_\text{ETVO}$ measurements (Section~\ref{sec:subjective}).
\item Through subjective analysis, we also demonstrate that both QoS and QoE methods are insufficient and also show where they fall short in characterizing TI sessions (Section~\ref{sec:subjective}).
\end{itemize}
%
%
\begin{figure}[!t]
    \centering
    \includegraphics[width=\columnwidth]{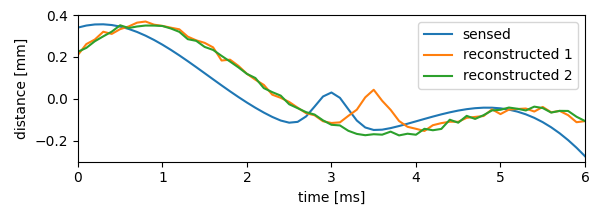}
    \caption{{\color{\additionColor}Illustration of the problem of RMSE when signals have time-offset. Two possible reconstruction signals -- 'reconstructed 1' and 'reconstructed 2' -- are shown along with the sensed signal. While the shape of 'reconstructed 1' is identical to sensed signal, it is delayed. On the other hand, 'reconstructed 2' misses the peak completely. However,  RMSE of 'reconstructed 1' turns out to be higher than that of 'reconstructed 2' due to insensitivity to time-offset.}}
    \label{fig:ti-blackbox}
    \vspace{-10pt}
\end{figure}

\section{Related Work}\label{sec:relatedwork}
\subsection{TI performance metrics}\label{subsec:performance}
In this section, we present an overview of the QoS and QoE metrics have been devised for evaluating TI systems. 

\subsubsection{\textbf{QoS}}
Several modular designs of TI systems use traditional QoS metrics, such as delay, jitter, packet loss, and throughput, for characterizing TI performance. While \textit{Admux}, an adaptive multiplexer for TI proposed by Eid et al. \cite{Eid2011}, uses all of the above metrics, the multiplexing scheme by Cizmeci et al. focuses on throughput and delay \cite{Cizmeci2017}. Hinterseer et al. \cite{Hinterseer2006} proposed a haptic codec that focuses on reducing the application throughput by transmitting only the perceptually significant samples. The congestion control scheme by  Gokhale et al. \cite{Gokhale2017} aims to contain delay and jitter within their permissible QoS limits. Further, a string of works has emerged recently that attempt to address the URLLC requirement of TI by leveraging the advancements in the field of 5G networks. The works in \cite{Li2018,sachs2018,kim2018} provide a detailed discussion on the vision and progress in this direction. While QoS metrics play a vital role in characterizing the network performance, they are insufficient for the comprehensive characterization of TI because they are signal-agnostic. In order to characterize the effects of a realistic network, one needs to understand how it affects the signal reconstruction and the resulting user experience. 

\subsubsection{\textbf{QoE}}
\label{sec:qoe}
Subjective QoE metrics aim to capture the quality of teleoperation by involving human subjects, typically 15-20, and have them subjectively grade their experience. A couple of works that adopt this approach include \cite{Basdogan2000, Yuan2014}. Since this method is cumbersome and resource-intensive, objective QoE metrics that estimate the quality of teleoperation as experienced by the human controller have also been designed.
Hinterseer et al.~\cite{hinterseer2005novel,hinterseer2008perception}, suggest a PD scheme for exploiting the idea that human perception has a logarithmic relationship with the haptic stimulus. A framework was developed to validate this by using the traditional Peak Signal-to-Noise Ratio (PSNR) of the reconstructed haptic signal as the QoE metric. Later, 
on the same lines, Sakr et al.~\cite{Sakr2007} introduced Haptic Perceptually Weighted Peak Signal to Noise Ratio (HPW-PSNR). 
Chaudhuri et al. \cite{Chaudhari2011} followed up on this to propose Perceptual Mean Square Error (PMSE) that maps MSE to the human perceptual domain. Recently, Hassen et al.~\cite{Hassen2018} proposed the Haptic Structure SIMilarity (HSSIM) index to improve the objective estimation of human perception. HSSIM extracts the similarity between original and reconstructed haptic signals. All of the above metrics are based primarily on RMSE, which suffers from a fundamental problem of not accounting for offset in the time domain. 
%

{\color{\additionColor}
\subsection{Generic similarity metrics}\label{subsec:similarity}
Determining the similarity between two signals is a classical signal processing problem and has been extensively researched due to its numerous applications, such as speech and gesture recognition~\cite{Salvador2007}. In this section, we discuss some of the techniques devised for this purpose and examine their applicability in extracting time and value-offset, which are crucial for TI applications.  
Cross-correlation computes the time-offset between the two signals that maximizes their dot product \cite{Rabiner1975}. It is well known that shared networks usually manifest highly non-deterministic and time-varying characteristics. Hence, a constant delay is an incorrect choice for representing the entire TI characteristics.
Another popular method known as Dynamic Time Warping (DTW) exists for signals encountering a time-varying delay \cite{Sakoe1978}. DTW conducts an exhaustive search to achieve sample-wise matching between the two signals in a manner that minimizes the cumulative Euclidean distance. It provides an extremely useful construct in determining \textit{how similar two signals are}. DTW functions as a practical starting point because it can already compare signals that differ in time. DTW is designed to find the similarity in sequences, for example, that two spoken words are the same, even when spoken at different speed and/or pitch. On the contrary, in teleoperation, the sensed and the reproduced signals are expected to be broadly similar. Hence, our problem is to find out \textit{how two similar signals are different}. 
While DTW completely solves its intended purpose, it is not designed for the stated objective of characterizing TI systems. Hence, we take DTW as the starting point in this work and perform substantial modifications to serve our purpose.

Several follow-up works on DTW exist, with each of them attempting to outperform DTW in one or more aspects. The most widely recognized ones include Edit Distance on Real sequences (EDR) \cite{Chen2005}, Edit distance with Real Penalty (ERP) \cite{Chen2004}, and Longest Common Sub-Sequence (LCSS) \cite{Vlachos2002}. However, they manifest the inherent characteristics of DTW and hence are unsuitable for TI, as will be detailed in the following sections. In the next section, we provide the necessary background of DTW as it forms the basis of the ETVO design.
}
\section{DTW: Background and analysis}\label{sec:limitations}
DTW measures the similarity between two sequences and is extremely useful for sequence classification problems like correlation power analysis, DNA classification, and notably, speech recognition. In this section, we present the necessary mathematical background for understanding the working of DTW. Next, we analyze to what extent DTW can characterize the performance of a TI session. We then construct a list of aspects that need to be addressed for DTW to fit our needs. 

\subsection{Mathematical Representation}
DTW constructs a \textit{warp path} that indicates a sample-wise mapping between two time-series that minimizes their cumulative Euclidean distance.
Let $\tilde{\bm{f}}$, $\tilde{\bm{g}} \subset \mathbb{R}^N$ denote the two $N$-length discrete time-series to be mapped. Let $\tilde{\bm{W}}$ denote the set of all possible warp paths between $\tilde{\bm{f}}$ and $\tilde{\bm{g}}$.
Let the $(k+1)$-th point of a warp path $\tilde{\bm{w}} \in \tilde{\bm{W}}$ be denoted as $\tilde{\bm{w}}(k) = (\tilde{\bm{w}}_0(k), \tilde{\bm{w}}_1(k))$, where $\tilde{\bm{w}}_0,\tilde{\bm{w}}_1\subset \mathbb{N}^{K}$ and $K \in [N, 2N-1]$. For example, the warp path in Figure~\ref{tkz:shortcommings} is given as [(0,0), (1,0), (2,0), (3,0), (4,1), (5,2), ...]. Essentially, $\tilde{\bm{w}}_0$ and $\tilde{\bm{w}}_1$ return the indices of $\tilde{\bm{f}}$ and $\tilde{\bm{g}}$, respectively.

The entries in $\tilde{\bm{w}}\in\tilde{\bm{W}}$ must meet the following conditions:
\begin{enumerate}
\item Monotonicity and continuity:
\begin{align*}
\tilde{\bm{w}}_0(k) \leq &\tilde{\bm{w}}_0(k+1) \leq \tilde{\bm{w}}_0(k)+1,\\
\tilde{\bm{w}}_1(k) \leq &\tilde{\bm{w}}_1(k+1) \leq \tilde{\bm{w}}_1(k)+1.
\end{align*}
\item Boundary:
\begin{align}\label{eq:boundary0}
\tilde{\bm{w}}(0)=(0,0), \tilde{\bm{w}}(K-1)=(N-1,N-1).
\end{align}
\end{enumerate}

The effect of these conditions is that subsequent samples are always put after their predecessors. 
DTW chooses the warp path that gives the minimum error ($l^2$-norm) between $\tilde{\bm{f}}$ and $\tilde{\bm{g}}$~\cite{berndt1994}. Hence, we get the error computed by DTW as  
\begin{align}\label{eq:DTW}
\text{DTW}(\tilde{\bm{f}},\tilde{\bm{g}})=\min_{\tilde{\bm{w}}\in\tilde{\bm{W}}} \sum_{k=0}^{K-1}\tilde{\delta}(\tilde{\bm{w}}(k)),
\end{align}
where 
\begin{equation}
\tilde{\delta}(\tilde{\bm{w}}(k))   (\tilde{\bm{f}}(\tilde{\bm{w}}_0(k))-\tilde{\bm{g}}(\tilde{\bm{w}}_1(k))^2.
\end{equation}

\noindent The computation of DTW$(\tilde{\bm{f}},\tilde{\bm{g}})$ is carried out as follows:
\begin{enumerate} 
\item Populate a cost matrix $\tilde{\bm{C}} \subset \mathbb{R}^{N\times N}$. Every point in this matrix gives a value indicating the cheapest path to that point from the start. Every element is given by,
\begin{multline}
\tilde{\bm{C}}[i,j] = \tilde{\bm{\delta}}(i,j)+\\\min 
\begin{cases}
\tilde{\bm{C}}[i,j-1],\\
\tilde{\bm{C}}[i-1,j-1],\\
\tilde{\bm{C}}[i-1,j]\},\\
\end{cases}
\forall i,j \in [0,N-1].
\label{equ:dtw_c}
\end{multline}

\item Backtrack from $\tilde{\bm{C}}(N-1,N-1)$ to $\tilde{\bm{C}}(0,0)$ to construct the warp path $\tilde{\bm{w}}$, where,
\begin{align*}
\tilde{\bm{w}}(\tilde{K}-1) = (N-1,N-1), \text{ and}
\vspace{-8pt}
\end{align*}
\end{enumerate} 
\begin{figure}[t]
\centering
\includegraphics[width=0.499\textwidth]{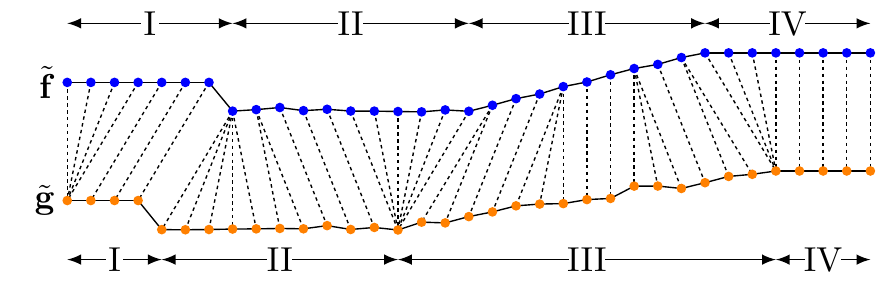}
\caption{Example of sample-wise alignment between signals $\tilde{\bm{f}}$ and $\tilde{\bm{g}}$ as per DTW. The dashed lines indicate the mapping between the samples.}
\label{tkz:shortcommings}
\vspace{-10pt}
\end{figure}
The time complexity of DTW is $O(N^2)$, although several algorithms for speeding up the computations exist~\cite{Salvador2007,Silva2016}.

\subsection{Challenges in applying DTW to TI}\label{sec:shortcomings}
We now investigate the characteristics of DTW that prevent it from being readily used for TI performance evaluation. In the context of TI, $\tilde{\bm{f}}$ and $\tilde{\bm{g}}$ represent the sensed and reconstructed signals, respectively.
\vspace{2mm}
\subsubsection{Boundary conditions cause unrealistic artifacts}\label{sec:problem_A}
The boundary conditions in Equation~\eqref{eq:boundary0} ensure that the extreme ends of the sequences are invariably aligned with each other. As a consequence, the delay is forced to be zero at the extreme ends. Segments 'I' and 'IV' in Figure~\ref{tkz:shortcommings} illustrate this. 
For TI applications, any non-zero delay systems will have a significant mismatch at the endpoints. This can be particularly significant when analyzing small sequences. 
\begin{figure}[t]
\tikzstyle{dot} = [draw,shape=circle,fill=black, scale =.3]
\centering
\begin{tikzpicture}[scale = 0.55, every node/.style={scale=0.7},rotate = 180, xscale = -1]

\def\H{0.25};

\node (input) at (0,0){};
\node (output) at ([shift={(0,1.5)}] input.center){};
\node [anchor=east] at (input){Input};
\node [anchor=east] at (output){Output};

\draw ([shift={(0.25,0)}] input.center) -- ++ (9.5,0);
\draw ([shift={(0.25,0)}] output.center) -- ++ (9.5,0);

\draw [fill=gray!20] ([shift={(3,-\H)}] input.center) rectangle ++(5,2*\H) node[pos=.5]{$N$};
\draw [fill=gray!20] ([shift={(4.5,-\H)}] output.center) rectangle ++(5,2*\H) node[pos=.5]{$N$};
\draw [fill=gray!20] ([shift={(1.375-0.5,-\H)}] input.center) rectangle ++(2,2*\H) node[pos=.5]{$M-1$};

\draw ([shift={(3,1.5*\H)}] input.center) -- ++ (0,1.5-3*\H);
\draw ([shift={(4.5,1.5*\H)}] input.center) -- ++ (0,1.5-3*\H);
\draw [>=latex, <->]([shift={(3,1.0)}] input.center) -- ++ (1.5,0);
\node (L) at ([shift={(3.75,1.0)}] input.center){};
\node [anchor=south] at (L.center){$\Delta T_{min}$};
\end{tikzpicture}
\caption{Illustration of extending the input sequence by $M-1$ samples in ETVO to avoid the start and end artifacts of DTW.}
\label{fig:lengths}
\vspace{-8pt}
\end{figure} 

\vspace{2mm}
\subsubsection{Unconstrained delay adjustments}\label{sec:problem_B}
The \textit{warp path} produced by DTW is generally not significant outside the algorithm. On the contrary, we intend to use the warp path as the estimated delay. 
As shown in Equation~\eqref{eq:DTW}, delay adjustments are no concern for DTW. 
In practice, this means that the warp path as a representation of time-varying delay is unrealistically erratic with many high-frequency changes that do not necessarily represent the TI system's behavior. For applications like speech recognition, high-frequency components in the warp path are of no consequence. However, this is unacceptable for TI, as delays and variations in delay are essential to understanding the performance.  Segments `II' and `III' in Figure~\ref{tkz:shortcommings} provide examples of significant and numerous shifts in delay that are disproportional to the differences in the signals. This can cause the TI system to appear to have a high variation in delay irrespective of the actual variation. 

\vspace{2mm}
DTW prefers to change delay when the velocity is as small as possible because that naturally lowers the l$^2$-norm. However, low-velocity regions do not correlate with when delay changes. Therefore the optimization criteria of DTW cause changes in the delay to not match in time with when the actual change occurs. Multiple examples can be found in Figure~\ref{tkz:shortcommings}. Segments `I' and `IV' start with an adjustment of delay. Despite that, the changes happen toward the end of the corresponding segments. At the start of Segment `II', there is a considerable delay change in a few samples before a small peak that causes the change.

In order to resolve the above issues and design suitable performance metrics for TI, we perform substantial refinements to DTW, as we describe in the next section.
\section{Design of TI Measurement Framework}\label{sec:etvo}
In this section, we present the mathematical foundation of the proposed framework for the characterization of TI sessions -- \textit{Effective Time- and Value-Offset (ETVO)}. Using this framework, we introduce two metrics: \textit{Effective Time-Offset} (ETO) and \textit{Effective Value-Offset} (EVO) to indicate the time- and value-offset, respectively, between the sensed and reconstructed signals. We use  \textit{effective} to indicate that the values show how the system appears to behave when considering it as a black box. For example, if a prediction method is used to make it seem like the signal is advanced by \SI{2}{ms}, ETVO should conclude that the delay is \SI{2}{ms} less.  Note that the unit of the value-offset matches the unit of the analyzed signals, which can be a position, velocity, force, and temperature, among others. 
\subsection{Proposed ETVO framework} \label{sec:change_A}
We now discuss our refinements for resolving the previously discussed issues of DTW for TI applications through the design of ETVO framework. 
\subsubsection{Relaxation of boundary conditions} \label{sec:change_B}
The first challenge we address is related to boundary conditions described in Section~\ref{sec:problem_A}. In order to change how the boundary conditions work, the mathematical structure is adjusted. Let $\bm{f}$ and $\bm{g}$ denote slices of the sensed and the reconstructed signals, respectively. For ease of explanation, we use the same notations as in DTW. However, we remove the accent (~$\tilde{}$~) to denote the ETVO counterparts. For DTW, the range of possible time-offsets depends on the length of the two time sequences. However, for ETVO, we define the possible time-offsets as a specified range. For TI systems, this is desirable because the range of expected time-offsets is caused by its components like the network and not the session length. The minimum time-offset is $\Delta T_{min}\in\mathbb{R}$, and the maximum time-offset is $\Delta T_{max} \equiv \Delta T_{min}+MT$, where $M\subset\mathbb{N}^+$, and $T$ is the sampling period of the signals. Let $N$ denote the length of $\bm{g}$. Every sample in $\bm{g}$ can take $M$ possible time-offset values. Hence, $\bm{f}$ should be of length $N+M-1$. If the first sample of $\bm{g}$ is located at $t=0$, then the first sample of $\bm{f}[k]$ should be located at $t=-\Delta T_{min}-(M-1)T$. This is illustrated in Figure~\ref{fig:lengths}.

DTW's warp path can be seen as a loose representation for time-offset. This concept is leveraged in ETVO to create a fine-grained representation of time-offset. Let $\bm{W}\subset\mathbb{N}^N$ denote the power set of possible warp paths to align $\bm{g}$ onto $\bm{f}$. The optimal warp path, the path that results in the smallest possible error, is denoted as $\bm{w}\in\bm{W}$, where $\bm{w}[k]$ indicates that $g[k]$ corresponds to $f[k-w[k]]$. Because ETVO limits its time-offsets to a specified range, the warp path represents relative time-offset directly. We denote ETO as the sample-wise time-offset corresponding to the alignment between $\bm{f}$ and $\bm{g}$. ETO can be derived directly from the \textit{warp path} and is expressed as
\begin{equation}
    \text{ETO}[k]=\Delta T_{min}+w[k].
\end{equation}
 We define the associated cost matrix as $\bm{C}\subset\mathbb{R}^{N\times M}$, where the $x$-axis indicates the sample index of $\bm{g}[k]$, and the $y$-axis is corresponding to time-offset and therefore ETO. Figure~\ref{tkz:algorithm} illustrates this concept, wherein the value at each entry of $\bm{C}$ indicates the cumulative cost of getting to that point. Specifically, the cost indicates $l^2$-norm of the most efficient warp path to get from the start of the signal to the current point. This is similar to the DTW counterpart $\tilde{\bm{C}}$ defined in Equation~\eqref{equ:dtw_c}, except that the $y$-axis here denotes the time-offset. The propagation through $\bm{C}$ is 
\begin{equation}
\bm{C}[i,j] = \delta(i,j)+\min 
\begin{cases}
\bm{C}[i-1,j],\\
\bm{C}[i-1,j-1],\\
\bm{C}[i,j+1],
\end{cases}
\end{equation}
where
\begin{multline}
    \delta(i,j) \equiv (\bm{g}[i]-\bm{f}[i-j+M-1])^2, \\\forall i \in [0,N-1], j \in [0,M-1].
\end{multline}
%
\begin{figure}[t]
\centering
\includegraphics[width=\columnwidth]{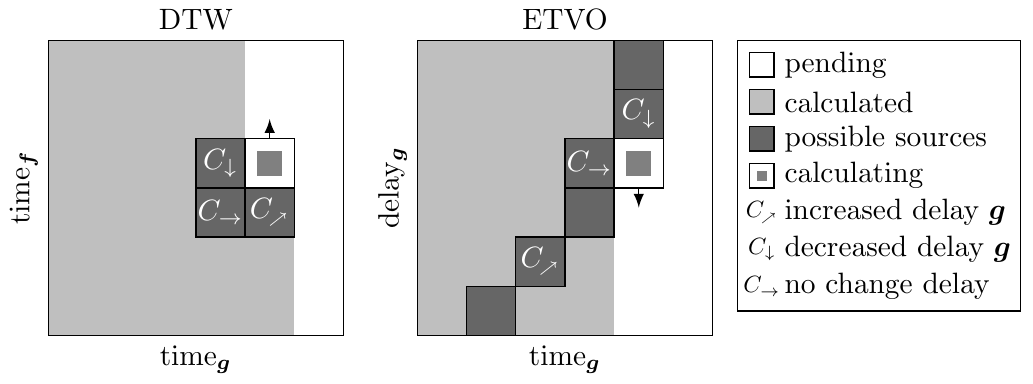}
\caption{{\color{\additionColor}Illustration of the population of $\bm{C}$ in both DTW and ETVO. Different types of changes in delay, indicated as
$C_\rightarrow$, $C_\downarrow$, $C\!\text{\tiny{$_\nearrow$}}$ are present in both the DTW and ETVO table to show their correspondence. A key difference between DTW and ETVO is that the latter also calculates multiple steps, increasing the possible sources as indicated with dark gray squares.
}}
\label{tkz:algorithm}
\vspace{-8pt}
\end{figure}
The three directions for calculating $\bm{C}$ correspond directly to the three directions in DTW as defined in Equation~\eqref{equ:dtw_c}. 
These new directions are indicated with $C\!\text{\tiny{$_\nearrow$}}$, $C_\downarrow$, and $C_\rightarrow$ indicating an increase, decrease, and no change in delay, respectively. 
An illustration of the resulting system and how the directions correlate between ETVO and DTW is shown in Figure~\ref{tkz:algorithm}. For this translated system, the monotonicity and continuity condition is given as
$$0 \leq \bm{w}(k+1)\leq \bm{w}(k)+1.$$
For DTW, swapping $\bm{f}$ and $\bm{g}$ leads to the same result. However, for the ETVO structure, the order of the signals is important. $\bm{g}$ projected onto $\bm{f}$ and $\bm{f}$ projected onto $\bm{g}$ would yield completely different results. 

An important effect of these changes is that it removes the boundary conditions enforcing the first and last sample of $\bm{f}$ and $\bm{g}$ to pair up. As a result, our framework now has the option to report non-zero delays for every sample in $\bm{g}$. The first column of $\bm{C}$ is initialized as $\bm{C}(0,*)= [0]^M.$ Every starting delay is assigned a zero cost. To remove the ending artifact, we let the last sample of ETO be chosen as the cheapest option, so that
\begin{align*}
\bm{C}(N-1,\bm{w}[N-1]) &\leq \bm{C}(N-1,j), &\forall j\in [0,M-1].
\end{align*}
As a consequence, not every sample of $\bm{f}$ has to be assigned a sample in $\bm{g}$. Therefore the DTW boundary condition given in Equation~\ref{eq:boundary0} is discarded for samples in $\bm{f}$. 

\subsubsection{Constraining delay adjustments} \label{sec:change_C}
In order to mitigate the issue of unconstrained delay adjustments in DTW (described in \ref{sec:problem_B}), we come up with substantial refinements to its design. For DTW, the warp path is designed as an intermediary, but for ETVO, we use the \text{warp path} as an indicator of the time-varying delay. First, let us define what a delay adjustment is in the context of ETVO. It is the change in estimated delay per unit time. $C_\downarrow$ and $C\!\text{\tiny{$_\nearrow$}}$ represent an increase and decrease in delay, respectively. A change in delay does not have to be of magnitude one but can be any positive integer. The dark gray squares in Figure~\ref{tkz:algorithm} indicate this.

In order to address the unconstrained delay adjustments, penalties are introduced to suppress adjustments that result in relatively minor improvements. We describe how multiple penalties are needed that target several aspects to achieve the intended result. 
We present the mathematical foundation behind the cost matrix $\bm{C}$ and describe the rationale behind the penalties.
\begin{align}
\phantom{C_\uparrow[i,j]}
&\begin{aligned}
\mathllap{C_\rightarrow[i,j]} &= \phantom{C}[i-1,j]
\label{eq:1}
\end{aligned}\\
&\begin{aligned}
\mathllap{C_\downarrow[i,j]} &= \min_{k\subset\mathbb{N}^+}\Big\{\bm{C}[i,j+k]\\
&\qquad\quad + \sum_{l=1}^{k-1}\delta(i,j+l) + kP_\text{prop}+P_\text{fixed}\Big\}
\label{eq:2}
\end{aligned}\\
&\begin{aligned}
\mathllap{C\!\text{\tiny{$_\nearrow$}}[i,j]} &= \min_{k\subset\mathbb{N}^+}\Big\{\bm{C}[i-k,j-k] + \sum_{l=1}^{k-1}\delta(i-l,j-l)\\
&\qquad\quad + kP_\text{prop}+P_\text{fixed}\Big\}.
\label{eq:3}
\end{aligned}
\end{align}
For every delay adjustment, we introduce two variables -- $P_\text{fixed}$ and $P_\text{prop}$, as shown in Equations~\eqref{eq:2} and \eqref{eq:3}. These correspond to a fixed penalty for every delay adjustment and a penalty proportional to the size of the delay adjustment, respectively. $P_\text{fixed}$ suppresses the number of delay adjustments, and $P_\text{prop}$ affects the magnitude of each adjustment. Together, these penalties suppress the delay adjustments estimated by the algorithm. The variable $P_\text{prop}$ balances between time and value-offsets. High penalties reduce the time-offsets and increase the value-offsets. ETVO performance approaches DTW when the penalties tend to zero. 
$P_\text{fixed}$ and $P_\text{prop}$ both reduce changes in time-offset at the expense of more value-offset, but with slightly different effects. $P_\text{prop}$ has a larger effect on the size of adjustments, while $P_\text{fixed}$ has a larger effect on the frequency of adjustments. The best candidate for each directions is calculated as shown in  Equation~\eqref{eq:1}-\eqref{eq:3} and is illustrated in Figure~\ref{tkz:algorithm}.
\begin{figure}[t]
\centering
\includegraphics[width=\columnwidth]{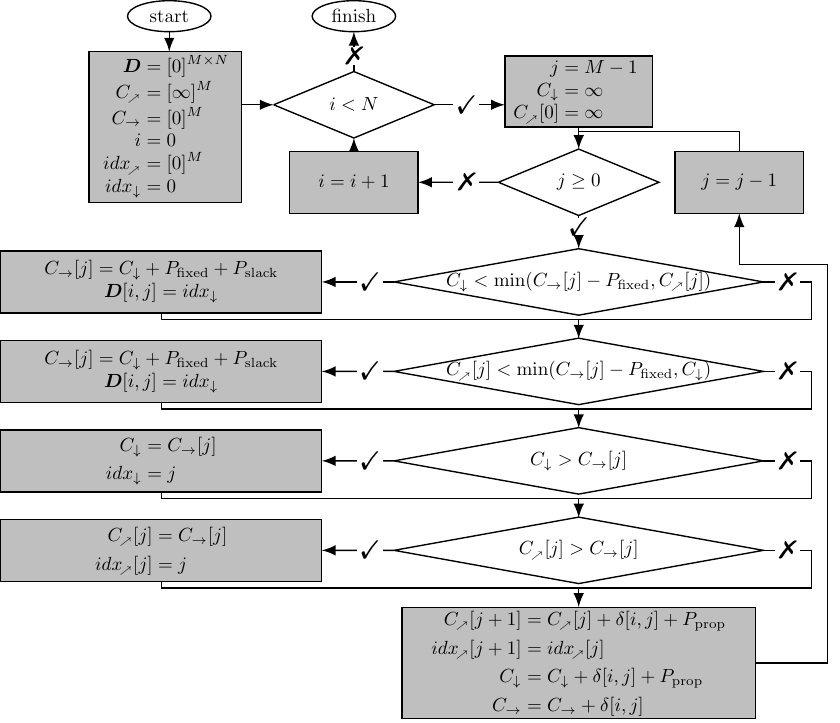}
\caption{Flowchart for finding the optimal way of traversing the delay, given the constraints specified for ETO.} 
\label{tkz:flowchart}
\vspace{-8pt}
\end{figure}

In the case of DTW, the delay adjustments do not have to align with the actual events that trigger the delay changes. It is beneficial for the algorithm to make changes when there is the least amount of velocity. The reason is that when the delay is adjusted, some samples are counted multiple times, and their contribution is less when the velocity is closer to zero. However, this tendency has little to do with when a change in delay actually occurs. For TI, such behavior makes analysis hard and makes the session quality estimation inaccurate. ETO should not be influenced by an event that occurs in the future. Note that $P_\text{fixed}$ and $P_\text{prop}$ do not address this issue of timing the delay adjustments. Therefore, we propose to introduce slack in delay adjustments where their timing is postponed until the slack penalty $P_\text{slack}$ is breached. $P_\text{slack}$ acts on top of $P_\text{fixed}$ and $P_\text{prop}$ for every delay adjustment, but is only added after an adjustment is made. The addition of $P_\text{slack}$ increases the likelihood that the delay adjustments match the events that cause them. With this, the overall cost matrix $\bm{C}$ is given as follows.
\begin{multline*}\label{eq:final_C}
\bm{C}[i,j] = \delta(i,j)\\
+\begin{cases}
\!C_\rightarrow[i,j] &\text{if } C_\rightarrow[i,j]\!<\!\min\!\Big\{\!C_\downarrow[i,j],C\!\text{\tiny{$_\nearrow$}}[i,j]\!\Big\},\!\!\\
\!C_\downarrow[i,j]\!+\!P_\text{slack} &\text{if } C_\downarrow[i,j]\!<\!\min\!\Big\{\!C_\rightarrow[i,j],C\!\text{\tiny{$_\nearrow$}}[i,j]\!\Big\},\!\!\\
\!C\!\text{\tiny{$_\nearrow$}}[i,j]\!+\!P_\text{slack} &\text{otherwise} \\
\end{cases}
\end{multline*}

\subsubsection{Defining EVO}
Unlike DTW, where the residual distance for every sample in the \textit{warp path} is aggregated into a single number similar to RMSE, we represent the value-offset as a time series that we call \emph{Effective value-offset} (EVO). Every sample of EVO indicates the error computed by $l^2$-norm from all samples of $\bm{g}$ compared to the corresponding sample in $\bm{f}$, excluding the penalties. When ETO increases or stays the same, only one sample of $\bm{g}$ is compared to $\bm{f}$. However, when ETO decreases, the EVO value for that sample is the $l^2$-norm between the output sample and several input samples.
This enables obtaining fine-grained information on how samples contribute to the value-offset. The mathematical description of EVO is given by 
\begin{multline*}
    \text{EVO}[i] = 
\begin{cases}
\sum_{l=\text{ETO}[k\!+\!1]}^{\text{ETO}[k]}\delta(i,l) &\text{if } \text{ETO}[i]\!>\!\text{ETO}[i\!+\!1],\\
\delta(i,\text{ETO}[i]) &\text{otherwise.}
\end{cases}
\end{multline*}
Due to this, there are spikes in EVO every time the ETO reduces by a large amount.
\subsubsection{Computational complexity}
\begin{figure}[]
\centering
\includegraphics[width=0.5\columnwidth]{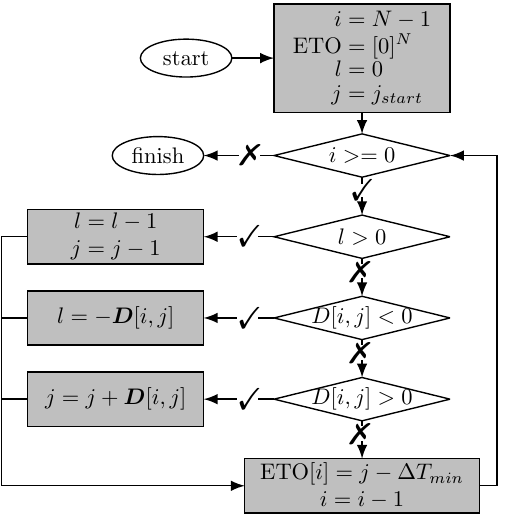}
\caption{Flowchart of the backtracking algorithm used to extract the ETO from direction matrix $\bm{D}$}
\label{tkz:flowchart2}
\vspace{-8pt}
\end{figure}

Besides presenting the ETVO framework, we also provide an efficient way of calculating ETO and EVO. The addition of $P_\text{fixed}$ results in a larger set of values to consider when finding the optimal path. Instead of the three adjacent locations, one has to consider a total of $M$ entries. Besides considering multiple entries, when backtracking to retrieve the delay, one must consider how many steps were taken. To store that information, we propose a direction matrix $\bm{D}\subset\mathbb{Z}^{M\times N}$. The number stored in $\bm{D}(k,i)$ indicates that the next point is at  $i+\bm{D}(k,i)$. Because the direction is stored, there is no need to store $\bm{C}$ entirely. There are three directions to consider: up, down, and forward. Each direction has one optimal source from which to start. If we remember only the optimal continuations in each stage of the algorithm, we only need to store a value and the corresponding index $C_\downarrow\!\subset\!\mathbb{R}$ and $idx_\downarrow\!\subset\!\mathbb{N}$ for downward propagation, arrays $\!C\!\text{\tiny{$_\nearrow$}}[i,j]\!\subset\!\mathbb{R}^M$ and $idx\!\text{\tiny{$_\nearrow$}}[i,j]\!\subset\!\mathbb{N}^M$ for upward propagation, and array $C_\rightarrow\!\subset\!\mathbb{R}^M$ for forward propagation. The resulting algorithm for populating $\bm{D}$ is illustrated with a flow chart in Figure~\ref{tkz:flowchart}. 

\begin{figure}
    \centering
    \includegraphics[width=\columnwidth]{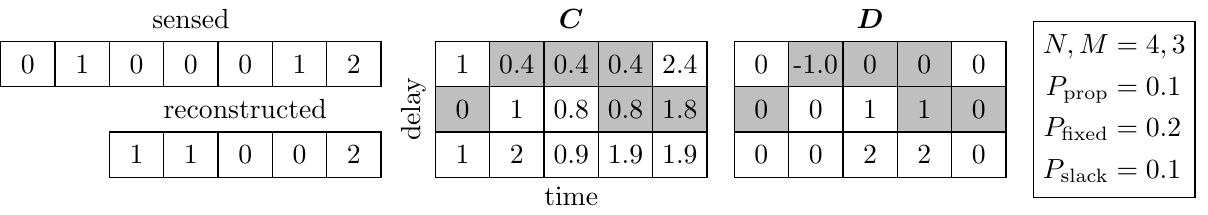}
    \caption{{\color{\additionColor}Numerical example of ETVO including the direction matrix. The gray cells indicate the optimal path chosen by ETVO.}}
    \label{fig:challenge}
\end{figure}
The backtracking algorithm needs to account for the multiple steps it can take. The calculation method matches how $\bm{C}$ was populated. Figure~\ref{tkz:flowchart2} shows a flow chart of the backtracking algorithm. The complexity of the algorithm for populating $\bm{D}$ is determined by two \textit{for loops}. One of the \textit{for loops} is looping through a fixed range $M$, which does not scale with signal length. Therefore the complexity is $\mathcal{O}(N)$. The complexity of the backtracking algorithm is bound by a single \textit{for loop}, so the upper bound on the combined set is also $\mathcal{O}(N)$. A numerical example of how $\bm{C}$ and $\bm{D}$ are populated is provided in Figure~\ref{fig:challenge}.

{\color{\additionColor}
\subsection{Quantitative Metrics for TI}
ETVO framework produces two time series -- ETO and EVO. While it is crucial to extract the fine-grained information about effective offsets for monitoring the performance in real-time and adapting the communication accordingly, it is also important to use them for performance benchmarking and comparing different TI solutions. Long-term averages serve this purpose better than time series. To this end, we propose two quantitative metrics that can be derived from ETO and EVO. 
\begin{enumerate}
    \item $T_\text{ETVO}$ -- the average end-to-end delay of ETO.
    \item $E_\text{ETVO}$ -- the average l$^2$-norm of EVO. 
\end{enumerate}
In this work, we use the above metrics for experimental evaluation of the effectiveness of ETVO in measuring TI performance. We intend to use ETO and EVO for TI performance monitoring and real-time adaptation in a future extension.   
}


%
\section{Testbed and objective analysis}\label{sec:setup}
To evaluate our proposed metrics, we develop a realistic TI testbed where a human user can interact with a remotely rendered virtual environment (VE) over a network. As a starting point for our testbed design, we consider a recently proposed testbed for simulating TI interaction~\cite{Bhardwaj2017}.
\begin{figure}[t]
\centering
\includegraphics[width=\columnwidth]{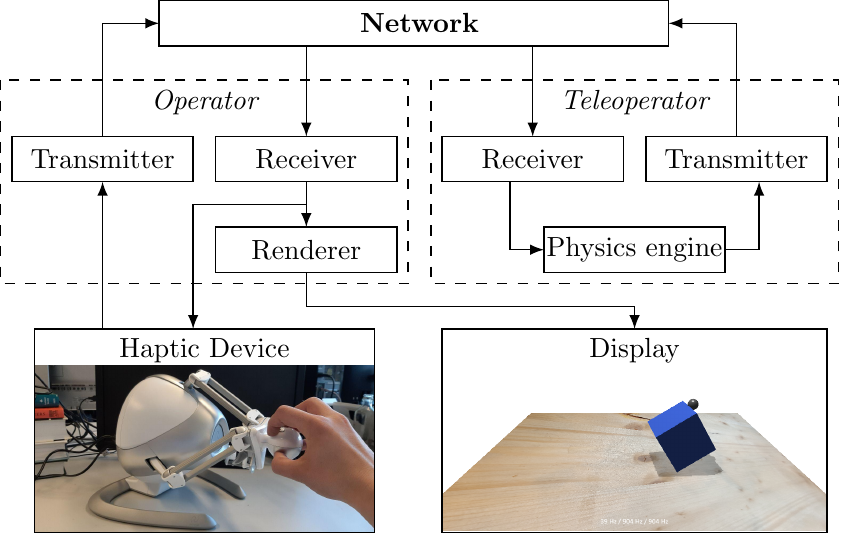}
\caption{A schematic overview of our experimental setup. The operator and teleoperator modules run on different computers that are not collocated. The physics engine resides in the controlled domain, resembling a real TI system using  Novint Falcon haptic device.} 
\label{fig:setup}
\vspace{-8pt}
\end{figure}
\begin{figure*}[t]
    \centering
    \includegraphics{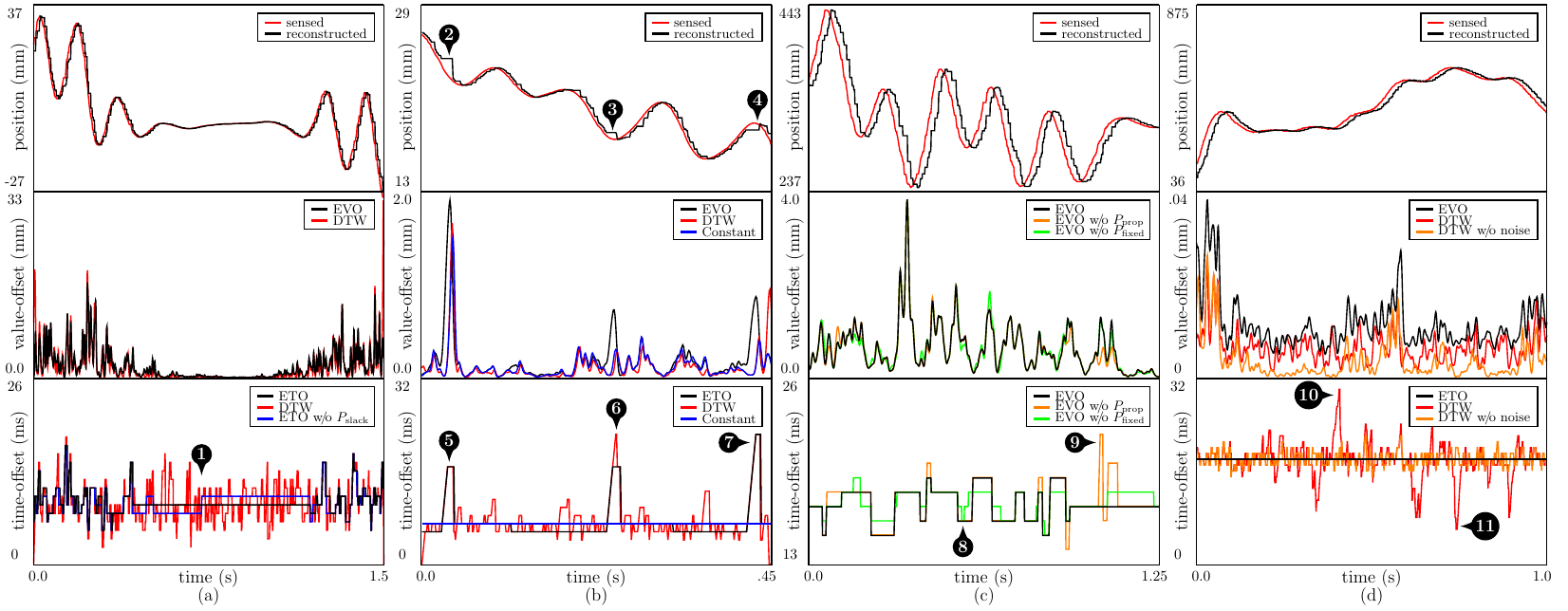}
\caption{Comparison of the performances of DTW and ETVO frameworks using a wide variety of experimental setups showing the  effects of (a)  $P_\text{slack}$, (b) uniform packet losses and perceptual deadband (PD) scheme, (c) $P_\text{prop}$ and $P_\text{fixed}$, and (d) addition of noise to sensed signal.}
    \label{fig:examples}
\end{figure*}
\subsection{Standard TI testbed} \label{subsec:stdsetup}
A TI testbed was proposed by Bhardwaj et al. \cite{Bhardwaj2017} and has been utilized to support haptic codecs standardization activities~\cite{steinbach2018haptic}. The testbed simulates a TI session by having the human participant interact with a VE via both haptic and visual feedback. The haptic device provides measurements at \SI{1}{kHz}, and the VE calculates force feedback at \SI{1}{kHz}. A visual rendering of VE is produced at a significantly slower rate of \SI{60}{Hz}. The haptic device used in this setup is a Novint Falcon \cite{novintfalcon}. Force calculation and visual rendering in the VE are implemented inside of Chai3D \cite{Chai3dfeatures}, where force calculations are performed by the \textit{Bullet real-time physics engine} \cite{bulletengine}.

Unfortunately, this testbed lacks the network component and is therefore not suited for mimicking TI interactions. Hence, we carry out significant refinements to the testbed in \cite{Bhardwaj2017} for realizing a networked TI testbed (described next).
%
\subsection{Networked TI testbed}\label{subsec:newsetup}
We extend the aforementioned testbed by decoupling the testbed into a master domain module and a controlled domain module, each residing on a different workstation, connected by a network. 
An overview of the entire system is shown in Figure \ref{fig:setup}. The master domain module senses the position of the haptic device. The controlled domain module houses the simulation of physics aspects. The physics simulation is a substitute for a TI application where the controlled domain module would house a real physical environment. The controlled domain module receives haptic device data through the feedforward channel and feeds it into the physics environment. Then the location and orientation of all dynamic objects and the calculated force are sent over the feedback channel to the master domain module. The master domain provides force feedback information to the haptic device based on data received from the feedback channel. The position and orientation values are used to displace all dynamic objects locally for the visual display. This overcomes the need for live video transmission, thereby saving significantly on the network bandwidth requirement. Further, this helps us to isolate the challenges of haptic feedback. 
We use data generated by our networked testbed to provide examples that demonstrate the efficacy of ETVO on a fine-grained scale. We also add white Gaussian noise to the sensed signals to evaluate ETVO's robustness to channel noise.

\textbf{Network emulation.} Netem, a standard network emulation tool, is used to emulate various network conditions, ensuring strong control over the network performance. This control is desirable, as the main purpose of the experiment is to analyze the performance of ETVO and not the testbed. Different network settings, such as delay, jitter, and loss, are introduced to evaluate the performance of ETVO. 
\subsection{Objective analysis}

The modifications to the basic DTW algorithm proposed in Section~\ref{sec:etvo} can be categorized into two groups. The first group deals with transforming the algorithm into an asymmetrical structure without start and end artifacts. The second group concerns the addition of penalties, which are required for improving the fine-grained analysis significantly. To illustrate these different aspects of ETVO, we picked four fragments from the haptic data trace.

We start by gauging the sensitivity of each of the schemes to the signal variations. 
In Figure~\ref{fig:examples}(a), it can be observed that at the extremes of the plot, ETVO shows fluctuations in time-offset estimation, but at areas with minimal changes, the frequency is reduced. This behavior reflects that delay will significantly impact the areas with extremes as opposed to the minimal areas. In contrast, DTW continuously fluctuates irrespective of the context. We also demonstrate the effect of $P_\text{slack}$ by comparing ETO with and without $P_\text{slack}$ (labelled as 'ETO w\!/\!o slack'). For the version without $P_\text{slack}$, it can be seen that the time offset changes in the minimal area  (as indicated with \blackcircled{1}). ETO with $P_\text{slack}$ postpones that decision to a more noticeable moment when the mismatch in delay leads to an observable difference. ETVO and DTW perform similarly in the value domain, despite the significantly higher number of delay adjustments performed by DTW. 
This example shows how ETVO makes evaluations that are context-aware. Further, note that DTW has a spike in value-offset on both edges because of the start and end artifacts. This behavior can be seen in the other examples as well.

In Figure~\ref{fig:examples}(b) there are periods of considerable value-offset due to a combination of bursty packet loss and PD. The Gilbert-Elliott model is used to induce bursty packet loss. There are three specific instances (markers \blackcircled{2} - \blackcircled{4}) where the combined effect of PD and bursty losses lead to a significant error in the reconstructed signal. In this case, DTW relentlessly adjusts the time-offset as the PD and losses are slightly degrading the signal. 
ETVO Chooses only to act when the effect is significant enough (markers \blackcircled{5} - \blackcircled{7}). 
The value-offset is smoothed with a Gaussian distribution for visual clarity. 

We now show the distinct effects of $P_\text{prop}$ and $P_\text{fixed}$, and demonstrate the importance of both. We show the results in Figure~\ref{fig:examples}(c), which has arrows with numbers that we will use as markers in this analysis.
We consider three different settings for algorithm parameters: \\\textcolor{white}{ii}(i) \textbf{[$P_\text{prop}$, $P_\text{fixed}$]} = [0.025, 0.05] (black curve), \\\textcolor{white}{i}(ii) \textbf{[$P_\text{prop}$, $P_\text{fixed}$]} = [0.05, 0] (amber curve), and \\(iii) \textbf{[$P_\text{prop}$, $P_\text{fixed}$]} = [0, 0.1] (green curve).

The values are chosen such that the overall strength of each setting is balanced but divided over $P_\text{prop}$ and $P_\text{fixed}$ differently to isolate the effect of omitting either of the penalties. Marker \blackcircled{8} indicates an event where scenario (ii) adjusts in a large number of small steps because there is no extra cost associated with using multiple steps. Marker \blackcircled{9} indicates an event where scenario (iii) causes a large step change but is limited in the number of steps because there is no extra cost associated with the size of a change. Scenario (i) has a similar performance in the value domain, but a significantly less cluttered ETO. 

Figure~\ref{fig:examples}(d) shows the effect that high-frequency noise has on DTW and ETVO. For this purpose, we add AWGN to the signal.
Both DTW and EVO are plotted with the noise added, while DTW w\!/\!o noise is a version of DTW without the added AWGN. High-frequency noise is a good example of a common way of signal distortion that DTW cannot deal with properly. Note that ETO outperforms the best case DTW, i.e. DTW w\!/\!o noise, demonstrating its noise resilience. Further, one can also notice the vulnerability of DTW to even a marginal amount of noise, causing time-offset to fluctuate vigorously.  

{\color{\additionColor}
\section{Subjective analysis}\label{sec:subjective}
Apart from the objective analysis, the networked testbed should provide a platform to facilitate subjective analysis. The setup is designed so that human operators can experience TI sessions and grade them based on subjective experience. We use this setup to demonstrate the efficacy of ETVO qualitatively. There are a few requirements for an experiment that benefit the statistical relevance of the test results.
\begin{inparaitem}
\item[\circled{1}] The participants should perform the same task multiple times under different settings.
\item[\circled{2}] To maximize the perception, the participants should concentrate. However, participants will have different levels of skill. Hence, the experiment must help the participants concentrate without placing high demands on their skill levels.
\item[\circled{3}] The task duration should be short and must enforce the operator to interact with the virtual environment continuously to generate haptic feedback. Long tasks can lead to fatigue, especially among older people. \end{inparaitem}

To meet the above requirements, we designed a \textit{target tracking} game that requires the participant to push a slider, labeled \textit{B} in Figure~\ref{fig:ingame}, left and right. During the test, the target (labeled \textit{A}) moves left and right. A participant has to push the slider to track the target as closely as possible. This task is consistent over multiple iterations, can challenge participants of any skill level, and because the slider has to move continuously, it invites continuous physics interactions. Hence all three of our requirements are met. 
\begin{figure}
    \centering
    \includegraphics[width=0.85\columnwidth]{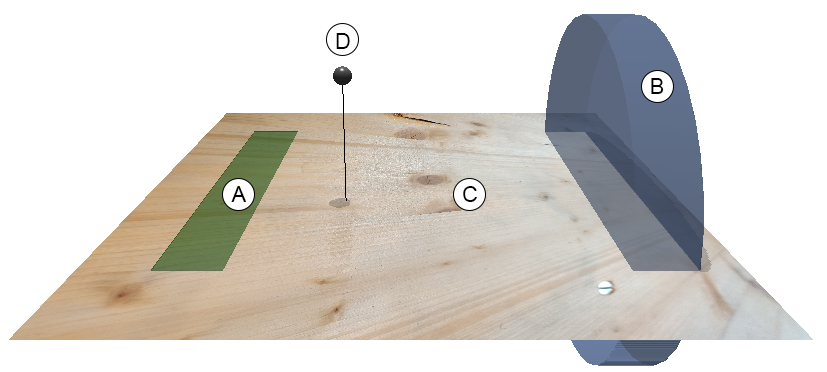}
    \caption{{\color{\additionColor}A snapshot of \textit{target tracking} game developed for the subjective performance evaluation of ETVO. `A' is the moving target that needs to be tracked by the slider indicated with `B'. `C' is a plane that serves as a rigid floor. `D' is the cursor that represents the position of the Novint Falcon in the virtual environment. A downward line and a shadow being cast on the plane to help the participant understand the location of 'D' better.}}
    \label{fig:ingame}
\end{figure}

\subsection{Network emulation}\label{sec:experiment_design}
During the experiments, users experience several instances of the same scenario while subjected to different emulated network settings as described in Section~\ref{subsec:newsetup}. To perform an extensive performance evaluation of ETVO, we consider a wide variety of network conditions, such as network delay, uniform loss (UL), and bursty loss (BL). We consider these settings in isolation and combinations. The bursty packet loss scenario is created using Netem's Gilbert-Elliot model. A bursty loss scalar $x$ is introduced, indicating the correlation between average packet loss $\pi_B$ and the probability of loss after a successful transmission $r$. Figure~\ref{fig:gemodel} shows how $x$ affects the Gilbert Elliot model. For satisfying the \SI{1}{kHz} haptic refresh rate, we apply linear extrapolation at the receiver. This takes care of irregular arrival of packets especially in cases of packet loss and PD scheme.
The way linear extrapolation is implemented is that in the master domain, the sensed position samples are used to estimate velocity, which is then included in each packet. This adds redundancy to the system that improves performance in most cases. 
For the subjective analysis experiments, $x=0.25$ was used.
\begin{figure}[t]
    \centering
    \usetikzlibrary{automata, positioning, arrows}
    \begin{tikzpicture}[scale=0.8, every node/.style={scale=1}]
    
    \node[state] (q1) at (0,0){$\pi_G$};
    \node[state, right of=q1] (q2) at (1,0){$\pi_B$};
    \draw [>=latex,->](q1) edge[above, bend left] node{$p=x\pi_B$} (q2);
    \draw [>=latex,<-](q1) edge[below, bend right] node{$r=x(1-\pi_B)$} (q2);
    \draw [>=latex](q1) edge[loop left] node{$1-p$} (q1);
    \draw [>=latex](q2) edge[loop right] node{$1-r$} (q2);
    \end{tikzpicture}
    \caption{{\color{\additionColor} Gilbert Elliot model and inclusion of scalar $x$ that allows one to change the distribution between bursty and uniform behavior without affecting the average packet loss. $\pi_G$ and $\pi_B$ are the average probability of successful and failed packet transmissions, respectively. $p$ and $r$ are the chance of switching states. }}
    \label{fig:gemodel}
\end{figure}
\subsection{Experimental Procedure}
Before the experiment, the participants are informed that the goal of the experiment is to investigate the effect of perceptual degradation. Each participant gets as much time as they want to familiarize themselves with the application with perfect network conditions, i.e. zero delay and zero loss. 
After that, a sequence of tasks, each lasting \SI{20}{sec}, is given, with a randomly chosen network setting per task. Participants grade the experience of each task on a scale of 10. An indication of how the user grades correlates with user opinions is shown in Table~\ref{tab:user_grade}. 
\begin{table}[]
    \centering
    \begin{tabular}{|c|c|}\hline
        10 & no perceivable impairment \\\hline
        8-9 & slight impairment but no disturbance \\\hline
        6-7 & perceivable impairment, slight disturbance \\\hline
        4-5 & significant impairment, disturbing \\\hline
        1-3 & extremely disturbing \\\hline
    \end{tabular}
    \vspace{2mm}
    \caption{{\color{\additionColor}Correlation between user grade and user opinion.}}
    \label{tab:user_grade}
\end{table}
\subsection{Participants}
The subjective study involved thirteen participants in the age group between 20 and 64 years, with an average of 30 years. Six participants were novice users of the haptic device. Nevertheless, every participant got ample time to familiarize themselves with the experimental setup. No participant suffered from known neurological disorders. Most of the data presented in this paper were collected during the COVID-19 pandemic. At all times, the safety regulations issued by the state were maintained, and extra care was taken to disinfect the equipment often. Because of these concerns, the number of participants is limited. This invites future research with more extensive data sets.
\begin{figure}
    \centering
    \includegraphics[width=0.51\textwidth]{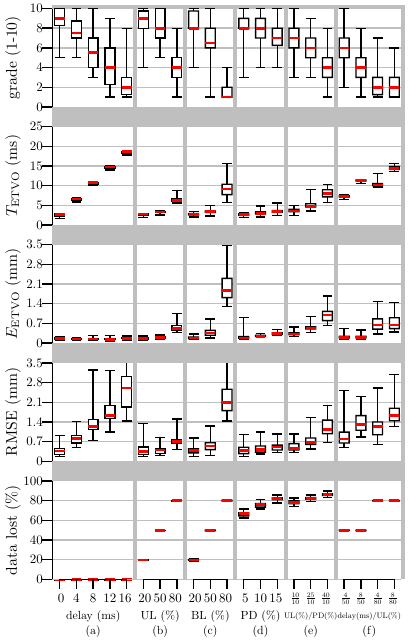}
    \caption{{\color{\additionColor}Demonstration of ETVO's strong correlation with the user grades along with comparison against QoS and QoE metrics. The experiments are performed under diverse settings of (a) constant network delay, (b) uniform random packet loss, (c) bursty packet loss, (d) perceptual deadband (PD) scheme, (e) uniform packet loss with PD parameter of 10\%, (f) constant delay with uniform packet loss. Other acronyms used: UL - uniform loss, BL - bursty loss. 
    }}
    \label{fig:results}
\end{figure}

\subsection{Performance analysis}
The data from all participants is aggregated and presented in Figure~\ref{fig:results}. The different types of network settings are separated by gray columns and the different measurements are separated by gray rows. 
The ETVO penalties are set to \textbf{[$P_\text{prop}$, $P_\text{fixed}$, $P_\text{slack}$] = [\SI{0.005}{}, \SI{0.01}, \SI{0.005}]}. We separately take up the performance comparison of ETVO with QoS and QoE methods. In all of our experiments, we employ linear extrapolation at the receiver, as described in Section~\ref{sec:experiment_design}.

\subsubsection{ETVO versus QoS methods}
In this section, we take up each network setting (described in Section~\ref{sec:experiment_design}) separately and shed light on the important observations. Each column in Figure~\ref{fig:results} corresponds to a different network setting. To substantiate the performance of ETVO, we also present discussions relating to different network settings.  

\vspace{2mm}
\noindent\textbf{1. Network delay.}
Figure~\ref{fig:results}(a) corresponds to the setting where we introduce a range of network delays.
As can be seen, $T_\text{ETVO}$ can track the network delay with negligible deviation. In addition, it also indicates an offset of approximately \SI{2.5}{ms}. This can be attributed to the discretization of haptic samples both at transmitter and receiver, OS-specific scheduling processes, and processing delay. Since ETVO considers the entire TI system as a black box, it is capable of extracting these local delays whose characterization would otherwise necessitate thorough system profiling. As is expected, the delay has a negative correlation with user grades, and  $T_\text{ETVO}$ reflects this accurately. Further, $E_\text{ETVO}$ correctly indicates negligible degradation in the value domain.  

\vspace{2mm}
\noindent\textbf{2. Uniform loss (UL).} In Figure~\ref{fig:results}(b), we introduce UL in the network. Before we move to discuss the performance of ETVO, we discuss an important concept that is crucial for interpreting our results.

The discretization of haptic signals inherently results in a time gap between haptic updates, which we call \textit{update duration}. This causes a lag between the master and controlled domains. which increases further when packets losses occur. In conventional networking applications, where latency constraints are far more relaxed, the update duration can be largely neglected. However, for TI systems this becomes significant. The average update duration, denoted by $\Delta t_\text{update}$, depends on the packet transmission rate and loss and can be expressed as
\begin{equation}\label{eq:avdelay}
\Delta t_\text{update} = 
\frac{1}{2f_s}+\frac{p}{f_s r(p+r)}, 
\end{equation}
where the first term is contributed by the sampling rate and the second by packet losses. $f_s$ is the rate at which the haptic device is sampled. 
We refer the reader to Appendix~\ref{sec:appendix} for a more detailed description and mathematical derivation of the average update duration. 

We apply this to Figure~\ref{fig:results}(b). Here, we have a packet transmission rate of \SI{1}{kHz}, and an average UL of 20\%, 50\%, and 80\%, resulting in $\Delta t_\text{update}$ of \SI{0.75}{ms}, \SI{1.5}{ms}, and \SI{4.5}{ms}, respectively. Note that in this setup, the network delay is zero. It can be seen that $T_\text{ETVO}$ computations corroborate well with the theoretical values accurately, in addition to the \SI{2.5}{ms} offset that we discussed previously. Further, the trend of $T_\text{ETVO}$ also matches that of user grade. On the other hand, QoS methods only measure only the packet loss present in the system without quantifying their effect on the user grades. 

$E_\text{ETVO}$ produces a similar trend as $T_\text{ETVO}$. A valid question is -- if the trend of $T_\text{ETVO}$ already matches the trend in the user grade, why do we need $E_\text{ETVO}$, or vice-versa? The answer to this can be found by comparing different network settings. If we compare the \SI{4}{ms} delay case in Figure~\ref{fig:results}(a) with \SI{80}{\%} UL in Figure~\ref{fig:results}(b), we see that the $T_\text{ETVO}$ is approximately equal. However, the corresponding user grades show a dramatic difference. Now, if we consider the information from $E_\text{ETVO}$ we can see that the latter case reports a significantly higher $E_\text{ETVO}$. This explains the lower user grade. This example highlights the significance of the combination of $T_\text{ETVO}$ and $E_\text{ETVO}$ being crucial for accurate estimation of TI performance. 

\vspace{2mm}
\noindent\textbf{3. Bursty loss (BL).}
In Figure~\ref{fig:results}(c), we present the results for the BL scenario. The average update duration introduced previously and expressed as Equation~\eqref{eq:avdelay} can be applied to the BL scenario also. However, the only difference compared to the UL scenario is the presence of a state-dependent aspect in BL. This means that whether the current packet is dropped depends on the state of the previous packets. Consequently, there is an increased chance of consecutive packet losses in the BL scenario than in the UL scenario. This dramatically increases the theoretical average update duration. 

Using Equation~\eqref{eq:avdelay} with $f_r=$ \SI{1}{kHz}, we obtain $\Delta t_\text{update}$ of \SI{1.5}{ms}, \SI{4.5}{ms}, and \SI{16.5}{ms} for BL of 20\%, 50\%, and 80\%, respectively. It can be clearly seen that $T_\text{ETVO}$ correctly reports a higher value than corresponding values of UL. However, as can be noticed in Figure~\ref{fig:results}(c), the theoretical worst-case delay is significantly higher than what is projected by $T_\text{ETVO}$. 
The reason for this is twofold.
Firstly, we use linear extrapolation in our experiments, while, for simplicity, we assumed a zero-order hold extrapolation in theoretical analysis. Linear extrapolation has a significantly higher impact for long episodes of packet loss. In some instances, the estimated velocity can even be higher than the sensed velocity, causing the linear extrapolation to lead the sensed signal. In this case, $T_\text{ETVO}$ is measured to be lower than the actual delay. On the other hand, linear extrapolation may also produce overshoot, values that might not exist in the sensed signal. This will be captured by $E_\text{ETVO}$ and not $T_\text{ETVO}$. 
Secondly, the ETVO penalties ensure that the time-offset is changed only when the value-offset reduces significantly. Because of this and the delay profile of bursty loss, the average delay as estimated by $T_\text{ETVO}$ drops significantly. This is further explained in Appendix~\ref{sec:sawtooth}.

\vspace{2mm}
\noindent\textbf{\underline{\smash{Observation on TI reliability.}}}
Note that the settings \SI{20}{\%} UL Figures~\ref{fig:results}(b), \SI{20}{\%} BL  (Figure\ref{fig:results}(c)) and the \SI{0}{ms} delay (Figure\ref{fig:results}(a)) have no significant difference in user grade. In case of UL even up to \SI{50}{\%} loss may become unnoticeable. This indicates that the user experience is not degraded even at significantly lower reliability. This highlights that the speculated ultra-reliability aspect of TI (\SI{99.9999}{\%}) needs thorough investigations going forward.

\vspace{2mm}
\noindent\textbf{4. Perceptual Deadband (PD).}
Next, we study the influence of the PD scheme without any packet loss in the network. As can be seen in Figure~\ref{fig:results}(d), the PD scheme dramatically reduces the number of transmitted packets. However, it is important to note that the PD scheme chooses to omit only the insignificant (redundant) data in the signal. Therefore, although the amount of packets received is significantly smaller, the user experience is good. It can be clearly seen that ETVO measurements match well with the user grades. Further, it can be seen that although the packets received in the case of PD of \SI{15}{\%} and UL of \SI{80}{\%} are similar, the user grade corresponding to the latter is substantially lower. While the packet reception rate is unable to identify this, ETVO is successful in capturing this aspect of the PD scheme. 


\vspace{2mm}
\noindent\textbf{5. Perceptual Deadband with uniform loss.}
We now include UL and PD schemes in conjunction. 
This scenario will see significant haptic updates being dropped by the network.
As can be expected, packet loss has a more detrimental effect on the user experience than a scenario without a PD scheme. This can be clearly observed in Figure~\ref{fig:results}(e). Even a \SI{20}{\%} UL with PD of \SI{10}{\%} results in noticeable change in user grade, whereas up to \SI{50}{\%} UL without PD scheme (Figure~\ref{fig:results}(b)) was barely perceivable. Indeed, ETVO can successfully capture this effect. Further, as per the packets received, the scenarios \SI{25}{\%} UL with PD of \SI{10}{\%} and \SI{80}{\%} UL without PD scheme (Figure~\ref{fig:results}(b)) behave in an identical manner. However, this contrasts with the user grade which is significantly lower in the former scenario. Once again, ETVO measures this accurately reporting higher $T_\text{ETVO}$ and $E_\text{ETVO}$ in the former scenario.  

\vspace{2mm}
\noindent\textbf{6. Network delay with uniform loss.}
In this setting, we use combinations of network delays (\SI{4}{ms}, \SI{8}{ms}) and UL (\SI{50}{\%}, \SI{80}{\%}). Figure~\ref{fig:results}(f) presents our findings of these scenarios. It can be seen that for a specific network delay, both $T_\text{ETVO}$ and $E_\text{ETVO}$ increase with UL. This is because with increasing UL, not only the update duration but also the value error increases. Further, for a specific UL, only $T_\text{ETVO}$ increases with network delay whereas $E_\text{ETVO}$ remains identical. This also makes sense as higher delay leads to degradation in the only time domain and not in the value domain. Interestingly, $T_\text{ETVO}$ does not accurately reflect the user grades specifically in case of (\SI{8}{ms}, \SI{50}{\%}) and (\SI{4}{ms}, \SI{80}{\%}). However, $E_\text{ETVO}$ in the latter case is significantly higher signifying yet again the importance of using both $T_\text{ETVO}$ and $E_\text{ETVO}$ in conjunction for measuring the TI performance. On the other hand, the packet reception rate misses out on all the fine details that govern the overall performance. This highlights the contribution of ETVO in measuring the TI performance accurately.

\subsubsection{ETVO versus QoE methods}
We now compare the performance of ETVO with QoE methods. As a representative of this broad category of metrics, we use RMSE, since, as described in Section~\ref{sec:qoe}, the vast majority of QoE solutions for TI are RMSE-based.
Hence, using RMSE helps us understand the fundamental limitations of these solutions. To reiterate, RMSE is oblivious of the degradation in the time domain when comparing the sensed and reconstructed signals.

We consider the same networks settings considered in the previous section. First, we consider the network delay only case in Figure~\ref{fig:results}(a). 
The RMSE measurements correspond to the position signal. Due to the inherent problem of RMSE,
the effect of delay is treated as value error, and therefore the misaligned samples are directly compared to each other. As a consequence, the calculated error term becomes heavily dependent on the velocity in the signal (speed of movement). For example, for a velocity of zero, a mismatch will not yield an error, but for a high velocity, a mismatch will yield a large error term. Certainly, more delay makes the system worse, but the dependency on velocity introduces a large variance into the performance estimation. This can be observed in Figure~\ref{fig:results}(a) in the RMSE row. On the other hand, ETVO treats the time-offset and value-offset separately, so that the correct samples are compared to each other, leading to significantly better performance. 

In Figure~\ref{fig:results}(f), there are combinations of delay and packet loss. 
For RMSE two observations can be made. Firstly, there is once again a high variance, that does not increase for higher packet loss. Secondly, the average RMSE has a similar trend to $T_\text{ETVO}$, but not the addition of $E_\text{ETVO}$. Thus, RMSE represents the average delay, with high variance, and this does not match the user grades. This illustrates the fundamental problem when not considering time mismatch. Due to this, samples are compared to the wrong counterpart, and therefore the shapes are incorrectly compared. These two examples illustrate the shortcomings of RMSE and by its extension all QoE methods that do not handle time mismatch. We also show how ETVO does handle mismatches and accurately reflects the user grades.

The problem of high variance in RMSE can also be observed in presence of packet losses, i.e. Figures~\ref{fig:results}(b)-Figures~\ref{fig:results}(e). In all of these cases, although the network delay is zero, the inherent system delay mentioned earlier is still present. As a consequence, RMSE is still subjected to a noticeable increase in variance. As opposed to this, even the small amount of delay is correctly reported by $T_\text{ETVO}$, and by its extension $E_\text{ETVO}$ is more accurate.
}

\section{Conclusion and future work}\label{sec:conclusions}
As the field of Tactile Internet (TI) is advancing fast, there is a strong need for quantifying its performance objectively. In this paper, we addressed the limitations of existing TI performance metrics. 
We found the Dynamic Time Warping (DTW) algorithm used in speech recognition as a suitable starting point. We highlighted certain issues in applying DTW directly to characterize the performance of TI applications. We developed a  an analytical framework -- \textit{Effective Time- and Value-Offset (ETVO)} for addressing them. 
Through objective analysis, using realistic TI experiments, we demonstrated the improvements of ETVO over DTW in terms of extracting fine-grained time and value offsets. 
{\color{\additionColor}
Through subjective analysis we identified the limitations of the existing QoS and QoE metrics that are used for TI systems. Further, under a wide variety of network settings, we show that ETVO measurements not only corroborate well with the user grades, but also outperforms QoS and QoE metrics. We derived an analytical expression for the average delay of TI sessions and showed that it matches well with ETVO measurements. Additionally, independent of ETVO analysis, we observed that even up to \SI{50}{\%} packet loss results in no significant reduction in user grades which contradicts with the anticipated ultra-reliability requirement prescribed for TI applications. We believe that this work will open up promising avenues for advancing the field of TI.  

While the current work looks at an offline session, we intend to solve the non-trivial problems of measuring TI performance objectively in real-time. Further, we would like to explore how different application-specific parameters can be adapted based on ETVO for maintaining the quality of TI sessions under time-varying network conditions.
}

\fi






\bibliographystyle{IEEEtran}
\bibliography{bibs}


\clearpage
{\color{\additionColor}
\begin{appendices}
\section{Derivation of update duration} 
\label{sec:appendix}

For ease of analysis, we split the time into steps of length equal to the sampling period i.e., $1/f_s$. Then we consider the possibility of packet loss for each time step. The time since the previous update is the same for every period after successful transmission, but for a sequence of consecutive packet losses, the time since the previous successful transmission continuously increases. Therefore, we use the Markov chain representation of the Gilbert-Elliot model given in Figure~\ref{fig:gemodel} by splitting up the packet lost state into multiple states to indicate how long ago a successful transmission occurred. The resulting Markov chain is shown in Figure~\ref{fig:modified_gemodel}.
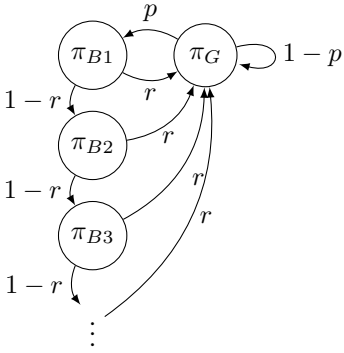
\begin{figure}[!h]
    \centering
\begin{tikzpicture}[scale = 1.0, every node/.style={scale=1}]
\def\H{1.2}

\node[state] (q0) at (1.5,0){$\pi_G$};
\node[state] (q1) at (0,-0*\H){$\pi_{B1}$};
\node[state] (q2) at (0,-1*\H){$\pi_{B2}$};
\node[state] (q3) at (0,-2*\H){$\pi_{B3}$};
\node[] (q4) at (0,-3*\H){\vdots};

\draw [>=latex](q0) edge[loop right] node{$1-p$} (q0);
\draw [>=latex,->](q0) edge[above, bend right] node{$p$} (q1);
\draw [>=latex,->](q1) edge[left, bend right] node{$1-r$} (q2);
\draw [>=latex,->](q2) edge[left, bend right] node{$1-r$} (q3);
\draw [>=latex,->](q3) edge[left, bend right] node{$1-r$} (q4);

\draw [>=latex,->](q1) edge[below, bend right] node{$r$} (q0);
\draw [>=latex,->](q2) edge[below, bend right] node{$r$} (q0);
\draw [>=latex,->](q3) edge[below right, bend right] node{$r$} (q0);
\draw [>=latex,->](q4) edge[right, bend right] node{$r$} (q0);
\end{tikzpicture}
    \caption{{\color{\additionColor}Markov chain of Gilbert-Elliot model with separate states for consecutive packet losses.}}
    \label{fig:modified_gemodel}
\end{figure}

Here, $\pi_{G}$ denotes the average probability of successful transmission, and $\pi_{Bn}$ is the average probability of previous successive $n$ packets lost.
Using this model we can derive the state vector $\pi$ and the transition matrix  $\bm{A}$. 
\begin{align}
\pi &= \begin{pmatrix}
\pi_G & \pi_{B1} & \pi_{B2} & \pi_{B3} & \cdots
\end{pmatrix}\\
\bm{A} &=
\begin{pmatrix}
1-p & p & 0   & 0   & 0&\cdots\\
r   & 0 & 1-r & 0   & 0&\cdots\\
r   & 0 & 0   & 1-r & 0&\cdots\\
r   & 0 & 0   & 0   &1-r &\cdots\\
\vdots &\vdots &\vdots &\vdots &\vdots &\ddots
\end{pmatrix} 
\end{align}

\noindent Using steady state distribution, we can now calculate the state by solving for 
\begin{equation}
    \pi \bm{A} = \pi
\end{equation}
By solving each of the resulting equations, we can formulate an infinite sum that describes the probability of each state as follows.
\begin{align*}
    \pi_G &= \frac{r}{p+r}\\
    \pi_{Bn} &= \frac{pr(1-r)^{n-1}}{p+r}\\
\end{align*}

We now know the probability of being in each state. The next step is to combine that with the average delay added by each state. Every state adds a delay equal to the total time spent till the previous update and half the period of a state, which is the inverse of the sampling rate $f_s$. Hence, $\Delta t_{update}$ is expressed as, 
\begin{align*}
    \Delta t_\text{update} &= \frac{1}{2f_s}+\frac{pr}{p+r}\sum_{k=1}^\infty \frac{k}{f_s}(1-r)^{k-1} \\
    \Delta t_\text{update} &= \frac{1}{2f_s}+\frac{p}{f_sr(p+r)}
\end{align*}
In the above equation, the first term represents the component due to the sampling frequency alone and the second term represents the impact of packet losses. 

\section{Difference between $T_{ETVO}$ and theoretical delay for bursty loss}\label{sec:sawtooth}
In this appendix, we provide insight into why the average delay as stated by $T_\text{ETVO}$ is lower than the theoretical estimate for network settings with bursty loss (BL). The reason has to do with the fundamental mechanics behind ETVO and its penalties. There is a cost associated with changing the delay, and it will only be done when the reduction in value-offset outweighs the penalties. 
\begin{figure}[!h]
    \centering
    \includegraphics[width=\columnwidth]{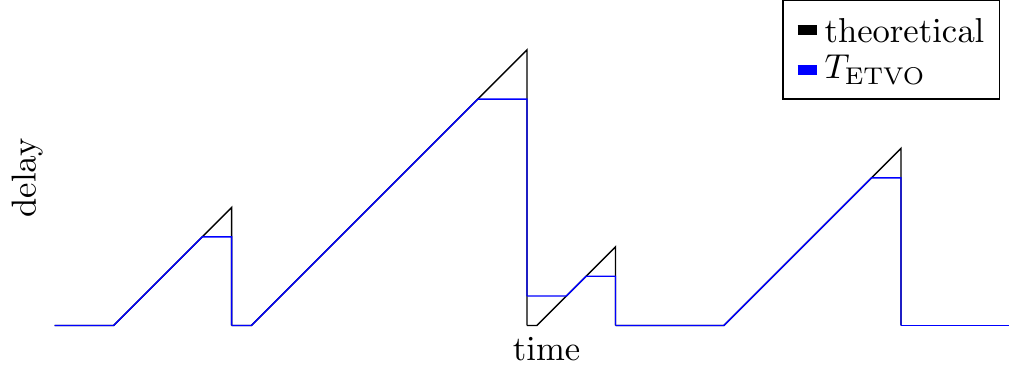}
    \caption{{\color{\additionColor}Illustration of delay profile for bursty loss}}
    \label{fig:sawtooth}
\end{figure}
Figure~\ref{fig:sawtooth} illustrates an example of what the theoretical delay for each sample looks like and how $T_\text{ETVO}$ interprets it. As you can see, the pattern created by bursty loss always has a narrow high delay part and frequently broader low delay part. If we consider that a cost needs to be paid to change the delay, there are only very few samples that are aligned for the high delay parts, so those few samples need to provide enough value-offset that it is worthy of. On the other hand, the lower parts are sometimes long, which means that adjusting the delay aligns with a considerable number of samples. In this case, it is more likely that the cost will pay for itself. Considering those observations, it is clear to see why the average delay as stated by $T_\text{ETVO}$ is lower than the theoretical calculation.
\end{appendices}
}


\end{document}